\documentclass[11pt,twoside]{article}

\usepackage{asp2006}
\usepackage{epsf}
\usepackage{graphicx}
\usepackage{psfig}
\usepackage{lscape}

\begin{document}
\title{Surprising Metallicity of a Newly Discovered M79 Post-AGB Star}   
\author{Timur \c{S}ahin,  David L. Lambert}   
\affil{Department of Astronomy and The W.J. McDonald Observatory, University of Texas, Austin, TX 78712, USA}    

\begin{abstract} 

A detailed chemical composition analysis based on a high-resolution ($R \simeq 35,000$) CCD spectrum
is presented for a newly discovered post-AGB star in the globular
cluster M79 for the first time. The elemental
abundance results of M79 Post-AGB star are found to be $[C/Fe]\simeq$-0.7, $[O/Fe]=$+1.4, $[\alpha
- process/Fe]\simeq$0.5, and $[s-process/Fe]\simeq$-0.1. The surprising result is that the iron abundance of the
star is apparently about 0.6 dex less than that of the cluster's red giants as reported
by published studies including a recent high-resolution spectroscopic analysis
by Carretta and colleagues.

\end{abstract}


\section{Introduction}   

\noindent In this study, published recently in full by \c{S}ahin \& Lambert (2009), we
report on an abundance analysis of the A-type m79 PAGB star  discovered by Siegel \&
Bond (2009, in preparation) in the globular cluster M79 and compare its composition to
that of the cluster's red giants. The initial mass of this star must have been
slightly in excess of the mass of stars now at the main sequence turn-off, say, $M
\simeq 0.8M_\odot$. The star's composition may be referenced to that of the cluster's
red giant stars for which abundance analyses have been reported. Comparison of
abundances for the PAGB and RGB stars may reveal changes imposed by the evolution
beyond the RGB; such changes are not necessarily attributable exclusively  to internal
nucleosynthesis and dredge-up. It was in the spirit of comparing the compositions of
the PAGB and RGB stars that we undertook our analysis. For the RGB stars, we use
results kindly provided in advance of publication by Carretta (2008, private
communication).  
\vskip 0.2 cm

 \begin{figure}
 \centering
\includegraphics[width = 20cm,,height=115mm, angle=90]{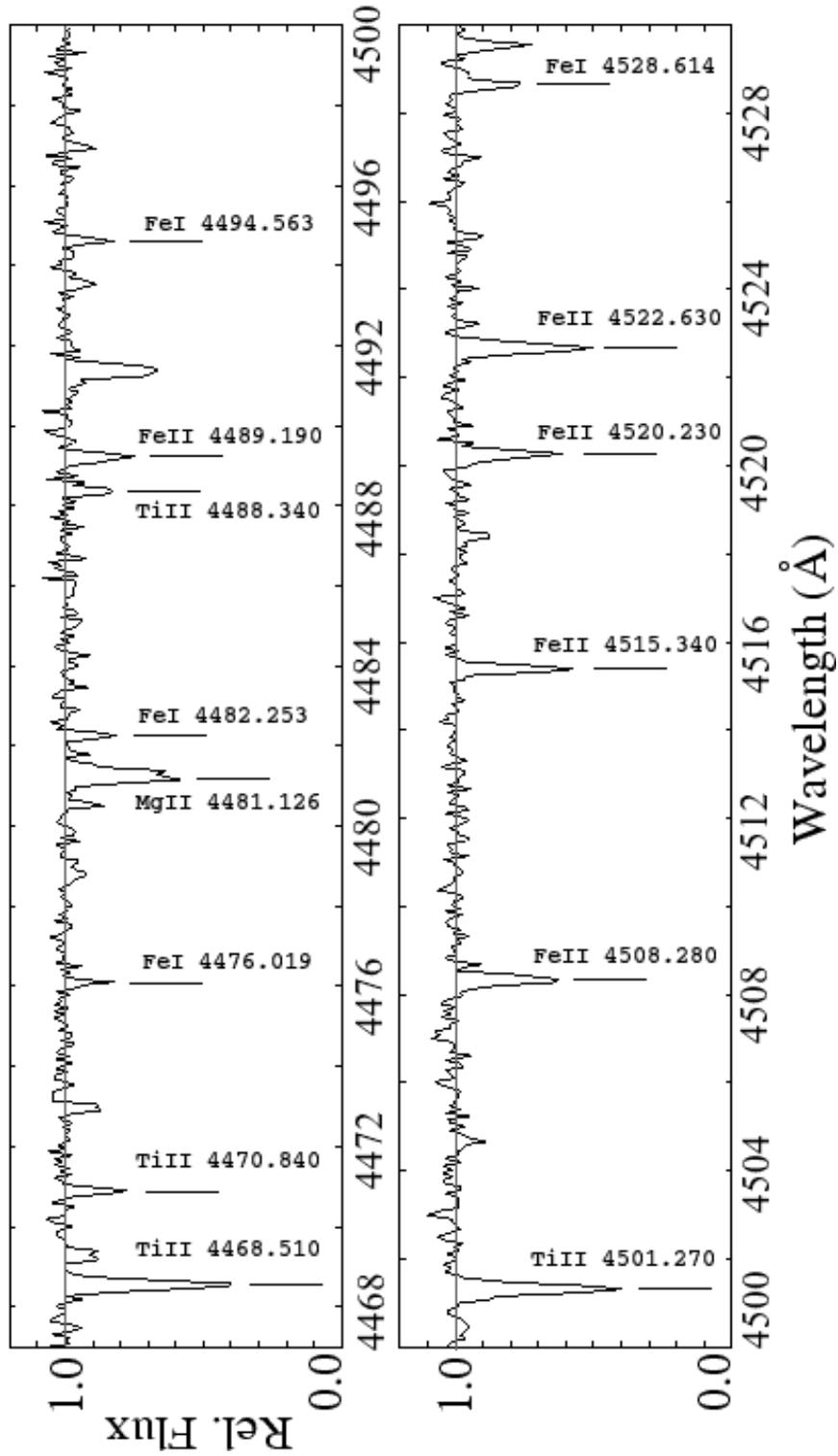} 
\caption{The spectrum for the PAGB star over the wavelength regions between
4467-4500 \AA\ (upper panel) and 4500-4530 \AA\ (lower panel). Selected lines
are identified.}
\end{figure}

\section{Observations}   

\noindent Spectra for the abundance analysis were obtained on five nights between 2008 January
15 and March 3 with the 2.7 meter Harlan J. Smith reflector and its $2dcoud$\'{e}
cross-dispersed \'{e}chelle spectrograph (Tull et al. 1995). Full spectral coverage is
provided from 3800 \AA\ to 5700 \AA\ with incomplete but substantial coverage beyond 5700 \AA
to 10\,200 \AA\ ; the effective short and long wavelength limits are set by the useful  S/N
ratio. A ThAr hollow cathode lamp provided the wavelength calibration. Flat-field and bias
exposures completed the calibration files. Observations were reduced using standard
procedures. A section of the final spectrum is shown in Figure 1. The heliocentric radial
velocity measured form the final spectrum is 211$\pm$5 km s$^{-1}$ with no evidence of a
variation greater than about $\pm$7 km s$^{-1}$ over the observing runs. This velocity is
consistent with the cluster's velocity of +207.5 km s$^{-1}$ given by Harris (1996). This
agreement between the PAGB star's velocity and that of the cluster confirms a result given by Siegel and bond (2009, in preparation).

 \begin{figure}
 \centering
\includegraphics[width=120mm,height=70mm,angle=0]{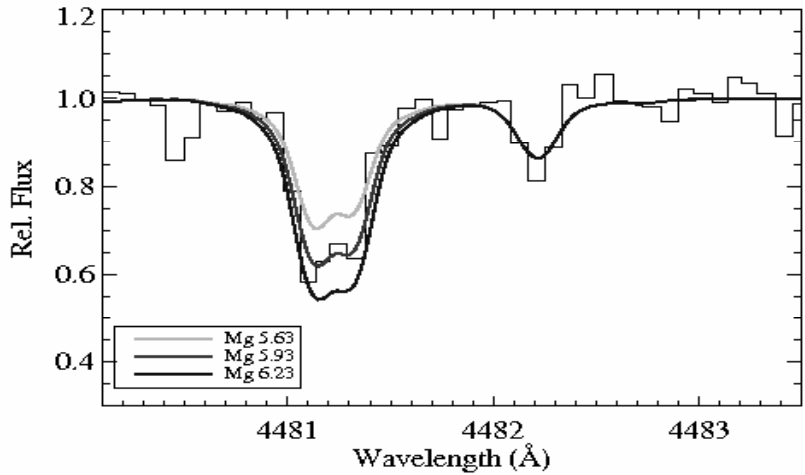} 
\caption{Observed and synthetic spectra around the
4481 \AA\ Mg\,{\sc ii} triplet lines.}
\end{figure}

\section{Spectral Analysis}   

\noindent The abundance analysis was undertaken with models drawn from the ATLAS9 grid (Kurucz
1993) and the line analysis programme MOOG (Sneden 2002). The models are line-blanketed
plane-parallel atmospheres in Local Thermodynamical Equilibrium (LTE) and hydrostatic
equilibrium with flux conversation. A model is defined by the parameter set; effective
temperature $T_{\rm eff}$, surface gravity $g$, chemical composition as represented by
metallicity $[Fe/H]$ and all models are computed for a microturbulence $\xi = 2$ km s$^{-1}$.
A model defined by the parameter set is fed to MOOG except that $\xi$ is determined from the
spectrum and not set to the canonical 2 km s$^{-1}$ assumed for the model atmosphere. In
\c{S}ahin \& Lambert (2009), we discuss several methods in an attempt to find consistent
values for the $T_{\rm eff}$ and $log\,g$ from photometry and spectroscopy. Application of
photometric and spectroscopic indicators of the atmospheric parameters for the PAGB star led
to the consensus choice of $T_{\rm eff}=6300$ K and $\log g$=0.8. A model with these
parameters (and a micro-turbulence $\xi=3.4$ km\,s$^{-1}$) fits not only the indicators but
also the locus in the $T_{\rm eff}$ versus $\log\,g$ plane provided by the constraint on the
star's luminosity and mass. Errors on these quantities are 300 K, 0.2 (cgs), and 0.5 km
s$^{-1}$ respectively. Synthetic spectrum fitting results for the 4481 \AA\ magnesium
triplet and resonance strontium lines at 4077 \AA\ and 4215 \AA\ are
presented in Figs. 2 and 3 as representative of $\alpha$- and $s$-process elements.

 \begin{figure}
 \centering
\includegraphics[width=120mm,height=70mm,angle=0]{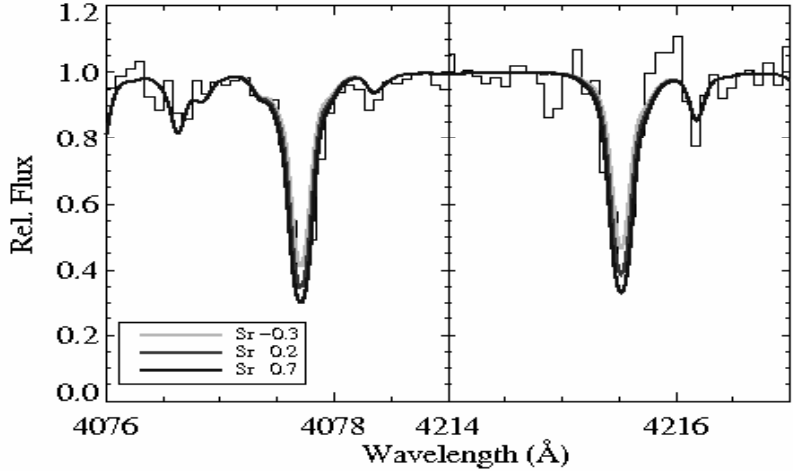} 
\caption{The observed spectrum around the Sr\,{\sc ii} 4077.7 \AA\ and 4215.5
 \AA\  resonance lines.}
\end{figure}

\noindent Table 1 summarizes the PAGB star's composition not only for the consensus model
but also for other three different model atmospheres and contrasts it with the mean
composition of the RGB stars. The abundance relative to iron [X/Fe] is also presented. The
standard notation is used here, i.e. $[X/Fe] = log(X/X_{\rm SUN} - log(Fe/Fe_{\rm SUN})$. The PAGB star's Fe abundance is $-0.5$ dex lower than that of
the RGB stars. For the majority of the investigated elements, the difference in abundance
$\log\epsilon$(X) in the sense (Ours $-$ Carretta) is within the range $-0.5\pm0.3$ dex, i.e.,
the differences are equal to $-0.5$ dex to within measurement uncertainties.The exceptions are
O, Na, Si, and Sr. A search (see
\c{S}ahin \& Lambert 2009) for an explanation of the composition difference between the PAGB
and RGB stars in terms of nucleosynthesis and dredge-up, dust-gas winnowing, and the first
ionization potential (FIP) effect proved negative.
\vskip 0.2 cm
 \begin{table}[ht]
 \small{
    \caption[]{Abundances of the observed species for M79 PAGB star are presented for four different model atmospheres. Also listed are
    abundances of the RGB stars in the same globular cluster analyzed by Carretta et al. (2008, private communication).}
       \label{}
   $$
       \begin{array}{@{}l||c@{}r@{}rrr||@{}c@{}}
          \hline
          \hline
               &            &       	   &  (T_{\rm eff}, \log\,g) &     &     &    \\
\cline{2-6}
 $Species$     &$RGB$       & $(6300,0.80)$ &$(6500,1.18)$  & $(6800,1.67)$   & $(7000,1.98)$ & \log\epsilon_{\odot}^{\star}   \\
\cline{2-6}
   &&  & \log\epsilon(X), $[X/Fe]$ &&   &    \\
          \hline
          \hline
 C$\,{\sc i}$       &  ..., ...  &\leq5.7,\leq-0.7&\leq5.7,\leq-0.8     &\leq5.9,\leq-0.8     &\leq6.0,\leq-0.9 & 8.39   \\
 O$\,{\sc i}$       & 7.29, +0.05&  8.0,+1.4  & 8.0,+1.2    & 8.0,+1.0     & 7.9,+0.8      & 8.66   \\
 Na$\,{\sc i}$      & 5.05, +0.42&\leq4.0 ,\leq-0.1&\leq3.8,\leq-0.5 &\leq3.8,\leq-0.7 &\leq3.7,\leq-0.9  & 6.17   \\
 Mg$\,{\sc i}$      & 6.13, +0.28&  5.8,+0.2  & 5.8,+0.2      & 6.0,+0.1    & 6.1,+0.1      & 7.53   \\
 Mg$\,{\sc ii}$     &  ..., ...  &  5.9,+0.4  & 6.0,+0.4      & 6.0,+0.1    & 6.0,+0.0      & 7.53   \\ 
 Si$\,{\sc ii}$^{\star}& 6.24, +0.29&6.2,+0.7 & 6.2,+0.6      & 6.2,+0.4    & 6.3,+0.3      & 7.51   \\
 Ca$\,{\sc i}$      & 4.93, +0.28&  4.7,+0.4  & 4.8,+0.4      & 4.9,+0.3    & 5.1,+0.3      & 6.31   \\
 Sc$\,{\sc ii}$     & 1.51, +0.04&  1.1,+0.1  & 1.2,+0.1      & 1.5,+0.1    & 1.6,+0.1      & 3.05    \\
 Ti$\,{\sc ii}$     & 3.49, +0.07&  3.2,+0.3  & 3.4,+0.4      & 3.6,+0.4    & 3.8,+0.4      & 4.90   \\
 Cr$\,{\sc i}$      & 3.96, -0.13&  3.5,-0.1  & 3.7,-0.1      & 3.9,-0.1    & 4.1,-0.0      & 5.64   \\
 Cr$\,{\sc ii}$     & 4.14, -0.02&  3.7,+0.1  & 3.8,+0.1      & 4.0,+0.1    & 4.1,-0.0      & 5.64   \\
 Mn$\,{\sc i}$      & 3.23, -0.53&\leq2.7,\leq-0.7&\leq 2.9,\leq-0.6 &\leq 3.1,\leq-0.6  &\leq 3.3,\leq-0.5  & 5.40   \\
 Fe$\,{\sc i}$      & 5.96, +0.00&  5.4,+0.0  & 5.6,+0.0      & 5.8,+0.0    & 5.9,+0.0      & 7.45   \\
 Fe$\,{\sc ii}$     & 5.94, +0.00&  5.4,+0.0  & 5.6,+0.0      & 5.8,+0.0    & 5.9,+0.0      & 7.45   \\
 Ni$\,{\sc i} $     & 4.54, -0.16&  4.3,+0.1  &  4.4,+0.1     & 4.7,+0.1    & 4.8,+0.2      & 6.23   \\
 Sr$\,{\sc ii}$     &  ..., ...  &  0.2,-0.7  & 0.4, -0.6     & 0.8,-0.5    & 1.0,-0.4      & 2.92   \\
 Y $\,{\sc ii}$     & 0.42, -0.27&\leq-0.3,\leq-0.5&\leq -0.1,\leq-0.4 &\leq 0.2,\leq-0.3    &\leq 0.4,\leq-0.3   & 2.21   \\
 Zr$\,{\sc ii}$     & 0.89, -0.16&\leq 0.7,\leq+0.1&\leq 0.9,\leq+0.2 &\leq 1.1,\leq+0.2 &\leq 1.4,\leq+0.4   & 2.59   \\
 Ba$\,{\sc ii}$     & 0.80, +0.13& 0.0,-0.1  & 0.2, -0.1     & 0.5,+0.0     & 0.8,+0.2      & 2.17   \\
 Eu$\,{\sc ii}$     &-0.62, +0.38& -1.0,+0.5  &-0.8, +0.6     &-0.5,+0.7    &-0.2, +0.9     & 0.52   \\
\hline
\hline
       \end{array}
   $$
}
 \end{table}

\section{Results}    

\noindent Many determinations of the metallicity [Fe/H] of cluster red giants have given
estimates near [Fe/H]$=-1.6$ (Standard notation is used for quantities [X] where
[X]=$\log$(X)$_{\rm star}-\log$(X)$_\odot$). For example, Zinn \& West (1984) give
[Fe/H]$=-1.69$ and Kraft \& Ivans (2003) give [Fe/H]$=-1.64$. Recently from high-resolution
UVES FLAMES spectra Carretta and colleagues (2008, private communication) performed an
abundance analysis for 20 elements obtaining [Fe/H]$=-1.58$ for a sample of ten RGB stars. The
exploration through quantitative spectroscopy of the newly discovered PAGB star in the
globular cluster M79 has led to an unexpected and, therefore, fascinating result: the standard
LTE analysis of the star has resulted in a metallicity different from that of the RGB stars
analyzed also by standard LTE techniques by Carretta.  The consensus model of (6300,0.8)
provides a [Fe/H] of $-2.0$ but the RGB analysis gives a [Fe/H]  of $-1.5$.

\noindent The star does not show
$s$-process enhancement. Oxygen and $\alpha$- process elements are enhanced. Abundances relative to iron appear to be the
same for the post-AGB star and the red giants for the 15 common
elements. It is suggested that the explanation for the lower abundances
of the post-AGB star may be that its atmospheric structure differs
from that of a classical atmosphere; the temperature gradient may be
flatter than predicted by a classical atmosphere.

\acknowledgements 
This research has been supported in part by the grant F-634 from the Robert A.
Welch Foundation of Houston, Texas.

\end{document}